# Comparing Deep Learning Models for Multi-cell Classification in Liquid-based Cervical Cytology Images

**Sudhir Sornapudi, Ph.D. student[1], Gregory T. Brown, M.D., Ph.D.[2], Zhiyun Xue, Ph.D.[2], Rodney Long, M.A.[2], Lisa Allen, B.S., CT[3], Sameer Antani, Ph.D.[2]**
[1]Missouri University of Science and Technology, Rolla, MO, USA; [2]Lister Hill National Center for Biomedical Communications, U.S. National Library of Medicine, Bethesda, MD, USA; [3]Diagnostic Systems Women's Health and Cancer, Becton Dickinson and Company, Durham, NC, USA

**Abstract**

*Liquid-based cytology (LBC) is a reliable automated technique for the screening of Papanicolaou (Pap) smear data. It is an effective technique for collecting a majority of the cervical cells and aiding cytopathologists in locating abnormal cells. Most methods published in the research literature rely on accurate cell segmentation as a prior, which remains challenging due to a variety of factors, e.g., stain consistency, presence of clustered cells, etc. We propose a method for automatic classification of cervical slide images through generation of labeled cervical patch data and extracting deep hierarchical features by fine-tuning convolution neural networks, as well as a novel graph-based cell detection approach for cellular level evaluation. The results show that the proposed pipeline can classify images of both single cell and overlapping cells. The VGG-19 model is found to be the best at classifying the cervical cytology patch data with 95 % accuracy under precision-recall curve.*

**Introduction**

Cervical cancer is the second most common cancer in women living in under-developed regions. Nearly 570,000 new cases were recorded in 2018 and about 311,000 women died from cervical cancer worldwide[1]. In the United States (2019) it is estimated that about 13,170 cases will be diagnosed for invasive cervical cancer and about 4,250 women will die from cervical cancer[2]. Fortunately, cervical cancer can be treated successfully if detected at early stage. LBC[3] for Pap (Papanicolaou) test is the gold standard for cervical cancer screening and has significantly contributed to reducing mortality. However, manual examination for detecting abnormal cells in a cervical cytology slide is a tedious process even for an expert cytologist. Expedient secondary reviews are conducted in areas marked by the pathologist or cytotechnologist with an ink marker by hand. There is a need for automated and computer-assisted technique for fast and efficient screening.

Though, generally abnormal cells have a relatively higher nuclei to cytoplasm ratio within a cell body[4], it is very time consuming and requires significant training and expertise to manually locate these abnormal cells under a microscope. While there are some automated approaches such as Becton-Dickinson's FocalPoint[a] and Hologic's ThinPrep[b], both manual and automated cytology are challenged by the high variability in cell size, shape and color, and complex morphology due to overlapped or folded cells. In recent years, computer-assisted automatic approaches have shown promising results in cell classification[5,6]. The current deep learning era has vastly improved the performance and the classification accuracy in various biomedical applications. While some methods avoid pre-segmentation step, much work in the literature shows considerable research in the direction of cell segmentation of cervical cytology images and single cell classification. Segmentation using superpixel-wise convolutional neutral network with dynamic shape modelling was employed by Tareef et al.[7] Watershed[8] and contour-seed pairs learning-based framework[9] were some of the successful approaches in segmenting overlapping cells. However, accurate segmentation of cervical cells is impeded by overlapped and clustered cells. Cell classification is the next step after segmentation. Pixel-level classification was employed by extracting traditional features and training an SVM classifier[10]. Similarly, block-level classification with SVM classifier was studied[11]. DeepPap[6] proposes CNN based cell classification on cell image patches, which are handpicked, cropped and centered at nuclei. All these approaches consider images containing only single cell or slightly overlapped cells, which are far from the real-world conditions. Traditional and machine learning methods were reviewed and discussed[12] for automated cervical screening from the pap-smear images. Our work lays foundation based on the stages involved in automated pap-smear analysis as mentioned in the literature review[12]. The stages for analysis typically comprise of image acquisition, preprocessing, identifying abnormal regions, feature extraction and finally classification.

---

[a] BD FocalPoint: https://www.bd.com/en-us/offerings/capabilities/cervical-cancer-screening/cytology-instruments---page-name-field/focalpoint-gs-imaging-system---page-name
[b] Hologic ThinPrep: https://www.hologic.com/hologic-products/diagnostic-solutions/aptimathinprep-cervical-health

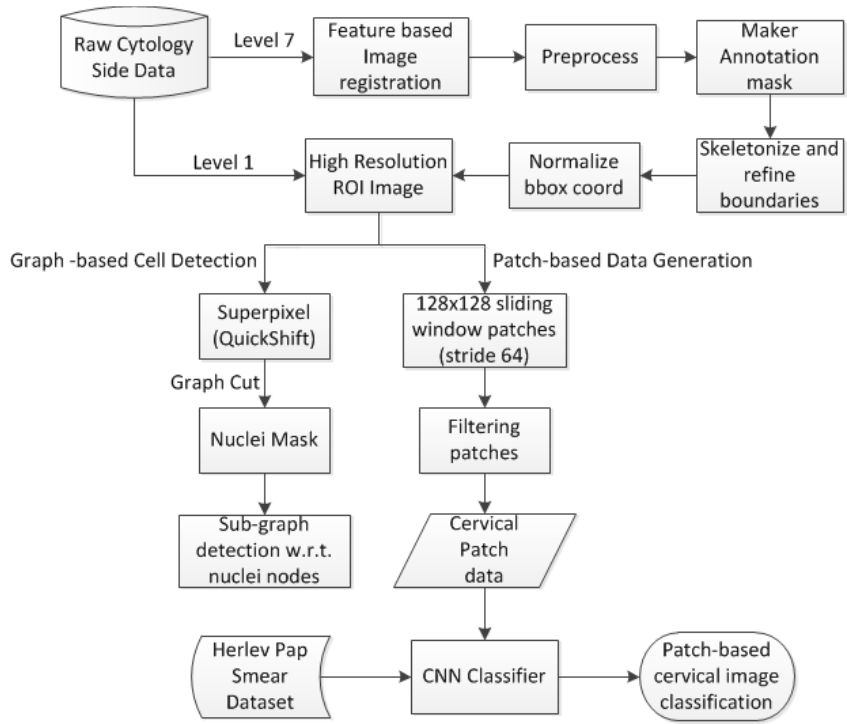

**Figure 1.** Framework of the proposed approach

The main objective of our study is to classify and locate the abnormal cervical cells. Understanding the challenges and limitations of the previous approaches, we aim to extract cell image patches from microscopic images under realistic conditions where cells may overlap; we evaluate different CNN models toward classifying the image patches.

**Materials and Methods**

The objective of the study is to generate clean labeled image data from the multi-level microscopic whole slide images to classify the cell image patches into normal and abnormal. First, raw cytology slide data (both clean and inked data) is preprocessed to obtain regions containing abnormal cells. The preprocessing comprises alignment of inked image data with respect to the clean image data through feature-based image registration. Lower resolution image levels (level 7) were used to speed up the process. Next, these lower resolution image levels were analyzed to obtain the ROI bounding box coordinates. This is accomplished by simple image processing techniques like image subtraction, thresholding, applying morphological operations and skeletonizing ink identified masks to detect and refine bounding boxes. Then, annotated image data is generated from the high resolution (level 1) images through accurately drawn abnormal cell boundaries on the ROIs. Finally, the labeled data was combined with Herlev Pap smear dataset[13] to train and test various CNN architectures and later evaluate the results to choose the best CNN model that classifies cytology data. A novel graph-based cell detection model was also proposed to identify the cell boundaries even under overlapping cases in the high-resolution image regions. The approach includes over segmenting the image with superpixels and connecting the centroids to create a graph and then identifying sub-graphs (cell regions) with respect to nuclei nodes obtained by applying appropriate graph cuts in the graph. The flow of this proposed approach is illustrated in Figure 1.

*Datasets*

The study uses two datasets. The first set comprises 25 cervical liquid-based cytology slides provided by Becton-Dickinson (BD) Corporation using their Sure Path technique[14]. In the case of abnormal slide data, there are a pair of slide images from each patient. One image contains only clean slide and the other image contains blue ink marks annotated by an expert cytotechnologist to indicate the regions containing abnormal cells (later, abnormal cells from these regions are identified by an expert pathologist). The ink marks can be easily cleaned with an alcohol swab. The glass slides were scanned both with the marks, and after their removal, using a Hammamatsu NanoZoomer 2.0-HT whole-slide scanner, producing digitized slides in a pyramidal tiled format with the file extension *ndpi*. The NDPI files resulting from the scanning are large in size (100's of megabytes of size data). The second dataset is the publicly available Herlev Pap Smear dataset (http://mde-lab.aegean.gr/downloads) where specimens are prepared via conventional Pap smear. The dataset contains 917 single cervical cell images. Table 1 shows the characteristics of the data used for the study.

**Table 1.** Characteristics of the datasets used for the study

| Dataset | Image type | Pixel Size (in μm) | #Normal | #Abnormal | File type | Total Images |
|---|---|---|---|---|---|---|
| BD Corp. Data | Whole slide image | 0.228x0.228 | 6 | 19 | NDPI | 25 |
| Herlev Data | Single cervical cell | 0.201x0.201 | 242 | 675 | BMP | 917 |

*ROI Detection*

While the entire cytology slide sample contains thousands of cells, just a few abnormal cells are sufficient indicators of abnormal screening. These abnormal cells may be surrounded by a large number of normal cells. Relative morphological, nucleic, or cytoplasmic appearance differences between cells helps pathologists to identify them. These regions are marked in ink on the glass slide. We use these markings on scanned slide image as region of interest (ROI) indicators. It is important to note that not all regions containing such abnormal cells were marked nor were individual abnormal cells identified within inked regions.

The NDPI format is organized as a pyramid structure with multiple levels of down-sampled subimages. A low-resolution image of level 7 is found suitable for easy preprocessing for detecting ROI bounding box coordinates. We correct any uniplanar misalignment problems between the blue ink-marked and ink-removed slides using image registration through feature based image alignment. The clean slide image is considered as a reference image, and stable ORB (Oriented FAST and Rotated BRIEF) feature points[16] along with descriptors are obtained. These key point features are mapped with the features in the inked slide image (target) as shown in Figure 2. Homography is calculated based on the mapping information using Random Sample Consensus (RANSAC) estimation technique[16]. The transformation is finally applied on the target inked slide image to map it to the reference clean image.

The preprocessing step also includes subtraction of blue (ROI ink) color space from red color space, which makes it easy to create a threshold ROI binary mask in lower resolution image. Skeletonizing and refining boundaries generated accurate ROIs. The coordinates of these ROIs were normalized and recorded. These ROIs are then cropped out from the high-resolution clean slide image (level 1) using the normalized bounding box coordinates data. Figure 3 shows the resultant intermediate output images.

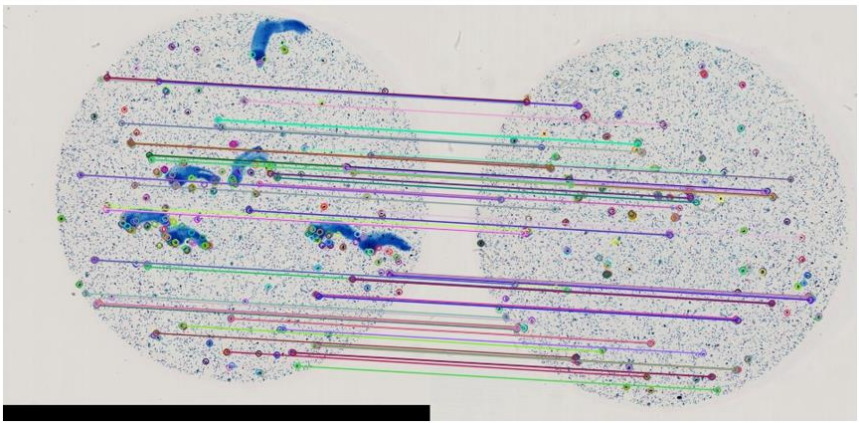

**Figure 2.** Matching keypoints in a low resolution image to align annotated slide image to the reference clean slide image

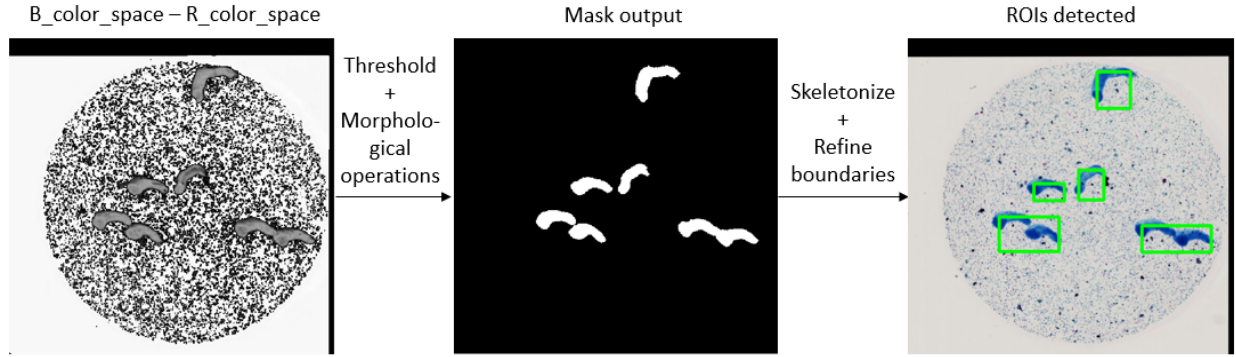

**Figure 3.** ROI detection from low resolution image

*Data Generation*

We propose two methods for cell data generation from the extracted ROIs: Graph-based cell detection and patch based data generation.

Graph-Based Cell Detection:

We chose this method to generate one cervical cell image (Herlev-like dataset). Initially the ROI image is over-segmented by generating superpixels through QuickShift approach[17]. The pixel intensities are averaged over each superpixel region. The resultant image is converted into a graph with centroid of superpixels as nodes, line connecting the adjacent nodes as edges and absolute difference of L2 normalized color intensities at the respective adjacent nodes as edge weight. Graph cut is used at an empirically determined threshold of 59. The resultant image is a binary mask for the nuclei present in the ROI image which form the graph nodes. The novelty of the proposed approach lies in generation of subgraphs out of this graph structure to detect cell body boundary. Figure 4 shows the intermediate image outputs. This approach could, in the future, automate segmentation of single cell images even where cells overlap.

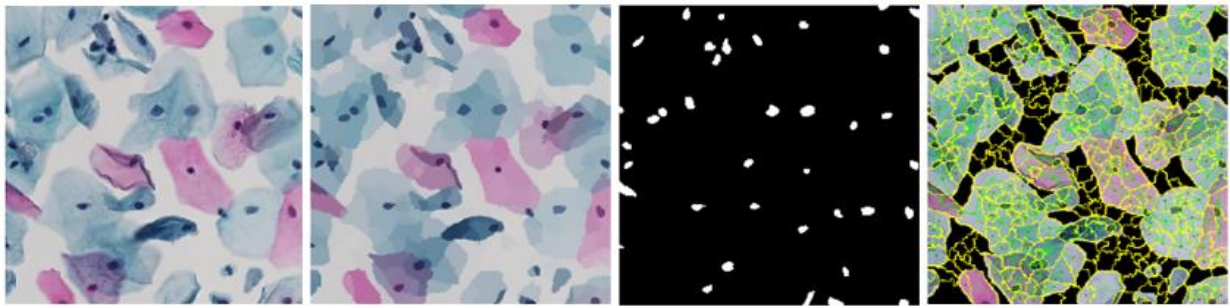

**Figure 4.** (from left to right) Original image; Averaging pixel intensities over superpixel regions; Nuclei mask; Graph structure connecting cellular regions

Patch Based Data Generation:

This approach is used to create images with real-world conditions from the high-resolution ROI regions where each image may contain multiple cervical cells along with overlaps. We use sliding window technique with stride 64 to create 128x128 patch images. The cells in the slide data are widely dispersed and contain more background. The patches containing more than 75% of background are discarded so that we get images with more cell information. The ground truth labels for the patch data are generated using the abnormal cell mask that is accurately and manually created with the help of an expert pathologist. Figure 5 shows how the abnormal cells were manually located. A patch image is labeled as abnormal if the object area in the abnormal cell mask is greater than 20% of the patch area (128x128). All the remaining patches are labeled as normal. Figure 6 shows examples of the final 128x128 labeled patch data used for the classification task.

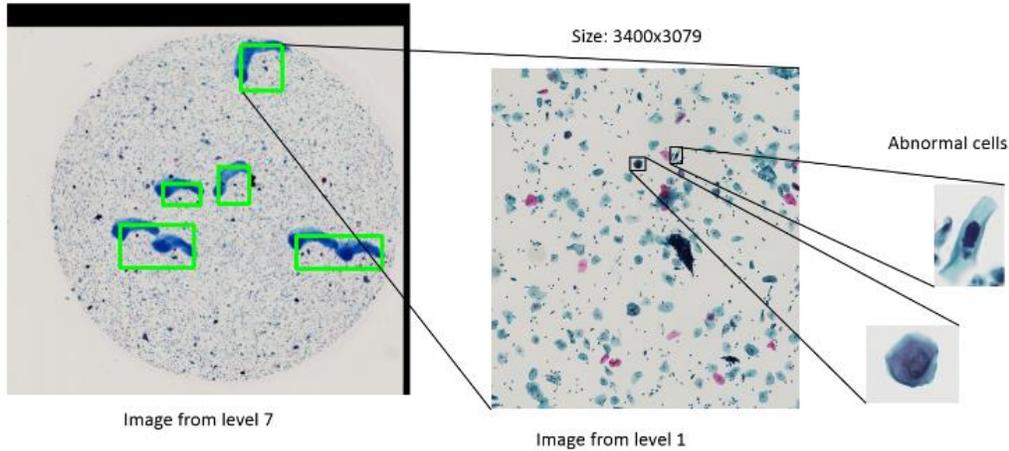

**Figure 5.** Manually locating abnormal cells in high resolution ROI regions.

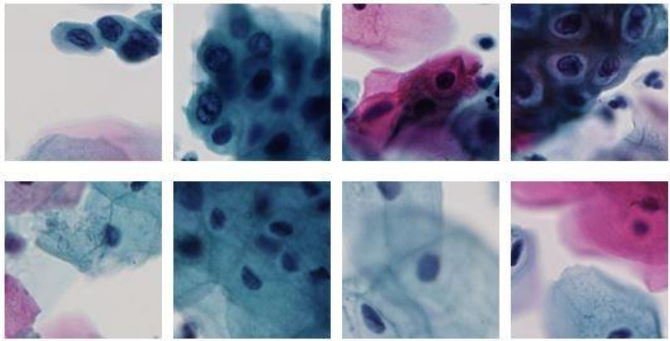

**Figure 6.** 128x128 patch image data from high resolution ROIs; Class labels:Abnormal (top-row) and Normal (bottom-row).

*Classification*

We consider images generated from patch based approach for the binary classification task. The patch data is completely randomized and split into train, validation and test data for training different CNN models. The abnormal patches are few in number compared to the normal patches. There were 2,060 abnormal cell patches generated, which are taken into the dataset and 2,060 normal cell patches were randomly selected to be a part of the dataset. The training and validation data sets consist of the entire Herlev Pap data and a part (75%) of the patches. The training and validation data split (65-10) with remainder of the patches forming the test data. This facilitates a balanced data distribution for image classification without any bias. Table 2 shows the data distribution among training, validation and testing datasets.

**Table 2.** Data split for training, validating and testing CNN classifier.

| Input for CNN Classifier | | Total | Patch data | Herlev data |
|---|---|---|---|---|
| Training | Normal | 1396 | 1200 | 196 |
| | Abnormal | 1760 | 1200 | 560 |
| Validation | Normal | 246 | 200 | 46 |
| | Abnormal | 315 | 200 | 115 |
| Testing | Normal | 660 | 660 | - |
| | Abnormal | 660 | 660 | - |
| Total | | 5037 | 4120 | 917 |

We used various well-established CNN classifiers in order to determine the best performing CNN model for the classification of cervical cytology patch data. Models VGG-19[18], ResNet-50[19], DenseNet-121[20], and Inception_v3[21] were fine-tuned whose weights are initialized with pre-trained ImageNet weights. In the training phase, all the layers are trained, and layer weights are updated for each epoch, and the model is trained with batch size of 32, learning rate 0.005 and momentum 0.9 (chosen empirically). This is a bi-classification task, so we employ cross entropy loss along with stochastic gradient descent optimizer. Each model is targeted to train for 500 epochs and ultimately the best weights are saved.

**Results and Discussion**

A total of 5,037 images (4,120 patch data + 917 Herlev data) were considered for the study. 1,200 patch data (660 normal, 660 abnormal) are utilized for testing and the CNN models are trained with remaining data. Pytorch deep learning platform is used to run the models on Nvidia DGX-1. The results tabulated are as shown in Table 3.

**Table 3.** Performance of various CNN models.

| Model | Confusion matrix $\begin{bmatrix} TN & FP \\ FN & TP \end{bmatrix}$ | ACC | PREC | REC | F1-Score | MCC |
|---|---|---|---|---|---|---|
| Resnet-50 | $\begin{bmatrix} 589 & 71 \\ 78 & 582 \end{bmatrix}$ | 0.8871 | 0.8913 | 0.8818 | 0.8865 | 0.7742 |
| VGG-19 | $\begin{bmatrix} 581 & 79 \\ 68 & 592 \end{bmatrix}$ | 0.8886 | 0.8823 | 0.8970 | 0.8896 | 0.7773 |
| Densenet-121 | $\begin{bmatrix} 611 & 49 \\ 131 & 529 \end{bmatrix}$ | 0.8636 | 0.9152 | 0.8015 | 0.8546 | 0.7329 |
| Inception_v3 | $\begin{bmatrix} 429 & 231 \\ 57 & 603 \end{bmatrix}$ | 0.7818 | 0.7230 | 0.9136 | 0.8072 | 0.5843 |

The results indicate that ResNet-50 and VGG-19 models performed better when compared to DenseNet-121 and Inception v3. DenseNet-121 has 0.9152 precision (PREC) value but has poor recall (REC) of 0.8015 with 131 false negatives. On the other hand, Inception_v3 has best recall (0.9136) compared to others, but poor at precision with 231 false positives. F1-score which presents balance between precision and recall is the best measure to evaluate the model performance. The F1-score of VGG-19 being 0.8896 votes VGG-19 as better model compared to ResNet-50 (0.8865). ResNet-50 is better at precision and VGG-19 is better at recall. In the field of biomedical image analysis higher recall value is always preferable, that is, lower false negatives are always recommended. Matthews correlation coefficient (MCC) provides much balanced measure by considering true and false positives and negatives. MCC for VGG-19 at 0.7773 makes it the best model compared to ResNet-50 (0.7742). Also, accuracy (ACC) wise VGG-19 is better than ResNet-50. The relatively small differences make both the models good competitors. To better understand the performance of the models at various classification thresholds, we plot the receiver operating characteristic (ROC) curve. Figure 7 shows the ROC plots for all the four models.

The best operating Q-point for the models in the cervical cytology image classification is the point where the curve has high sensitivity (true positive rate) and high specificity (1 – false positive rate). VGG-19 and ResNet-50 performed similarly with their best Q-points compared to other models. The accuracy under the ROC curve (AUC) was found to be 0.95 for both the models.

ResNet-50 and VGG-19 are good at generalizing the cervical cytology data with the pre-initialized ImageNet weights. VGG-19 is a shallow network compared to models under study and uses 3x3 convolutional layers stacked up along the depth of the network has proven to be a better model for the cervical cytology classification task in our study. ResNet-50 being a 50-layer deeper network, addresses its vanishing gradient problem through their residual learning blocks. This makes it a good competitor against VGG-19. Each layer in DenseNet-121 is fed with the outputs from the previous layers which improve the feature propagation and alleviate the vanishing gradient problem. Although DenseNet-121 is more efficient on some image classification tasks, it could not outperform the VGG-19 model on the cytology image classification. The 42-layer deep Inception v3 model factorizes convolutions and aggressively reduces dimensions which reduces the computational cost but could not maintain the quality in classifying the cytology image data.

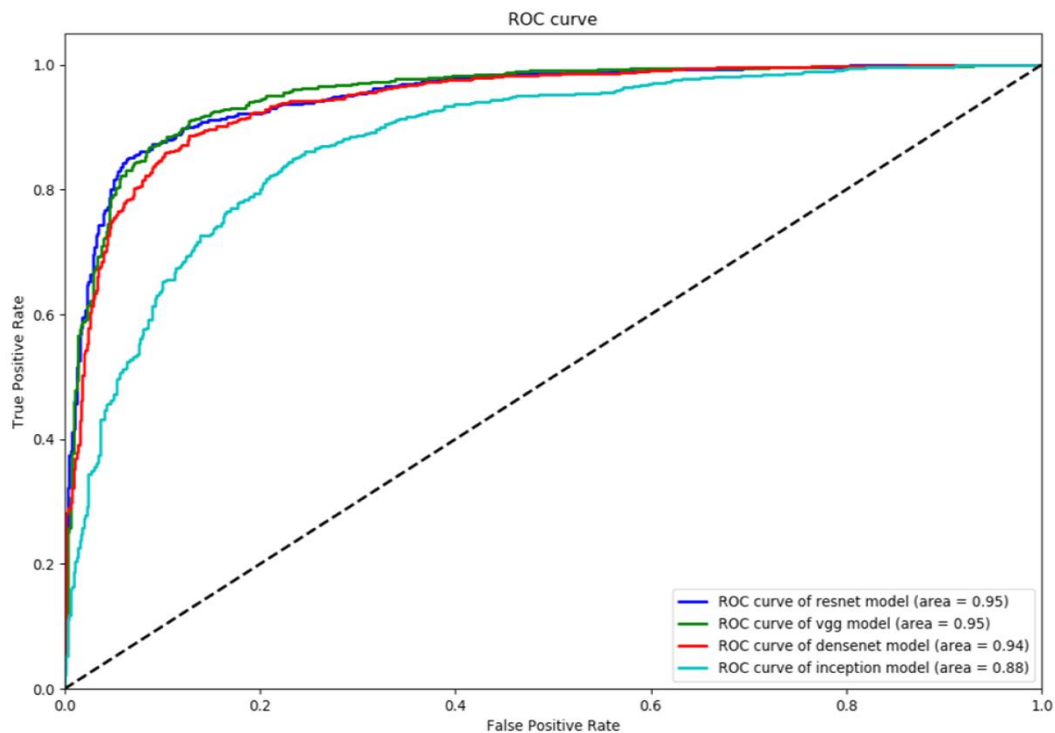

**Figure 7.** ROC curve of the four CNN models

We have further explored the performance of our best model, VGG-19, on the 128x128 abnormal cell patch data extracted with respect to the centroid of each object in the abnormal mask. 203 abnormal image patches were obtained and tested with VGG-19 model. We achieved an accuracy of 0.8778 upon prediction on these 203 abnormal image patches. This indicates that the VGG-19 model is better at classifying the cytology cell images, even under challenging cell-overlapping conditions.

While this work is preliminary, it demonstrates the capability of deep learning to recognize abnormal cells in cervical cytology specimens. With further refinement this work could be incorporated into a production level tool to assist pathologists with pre-screening and quality assurance, thereby improving a pathologist's efficiency and accuracy.

**Conclusion**

We have successfully developed and evaluated a prototype pipeline for the classification of cervical cytology slide images. The process automatically generated cleaner labeled patch image data for training and testing convolution neural networks. Our approach considers realistic conditions of overlapping cells which is superior to state-of-the-art classification techniques that rely on segmented cells. We investigated various CNN models for successful classification of cytology image data, and found VGG19 and ResNet-50 were similar best performers with our data. A novel graph-based cell detection technique was also proposed which may be used for developing cell analysis techniques. Our work represents a novel approach for classifying cytopathology image data, using real-world samples.

**Acknowledgment**

This research is supported by the Intramural Research Program of the National Institutes of Health, National Library of Medicine, and Lister Hill National Center for Biomedical Communications. In addition, we thank Becton Dickinson Diagnostic Services for de-identified SurePath liquid cytology Pap slides and Stephen Hewitt, MD, PhD in the Laboratory of Pathology, National Cancer Institute for assistance with whole-slide scanning.